\documentclass{ws-procs9x6}

\newcommand{\AmS}{{\protect\the\textfont2
  A\kern-.1667em\lower.5ex\hbox{M}\kern-.125emS}}

\def \znbb {$0\nu\beta\beta$}

\newcommand{\ba}[1]{\begin{eqnarray} \label{(#1)}}
\newcommand{\ea}{\end{eqnarray}}

\begin{document} 

\title{First Evidence for Neutrinoless Double Beta Decay}

\author{H. V. KLAPDOR-KLEINGROTHAUS}

\address{Max-Planck-Institut f\"ur Kernphysik,\\ 
P.O. Box 10 39 80, D-69029 Heidelberg, Germany\\ 
Spokesman of HEIDELBERG-MOSCOW and GENIUS Collaborations\\
E-mail: klapdor@gustav.mpi-hd.mpg,\\
 Home-page: http://www.mpi-hd.mpg.de.non\_acc/}

\maketitle


\abstracts{
	Double beta decay is indispensable to solve the question of 
	the neutrino mass matrix together with $\nu$ oscillation experiments. 
	Recent analysis of the 
	most sensitive experiment since nine years - 
	the HEIDELBERG-MOSCOW experiment in Gran-Sasso - 
	yields a first indication for the neutrinoless decay mode. 
	This result is 
	the first evidence for lepton number violation and proves 
	the neutrino to be a Majorana particle. 
	We give the present status of the analysis 
	in this report.
	It excludes several of the neutrino mass scenarios 
	allowed from present neutrino oscillation experiments - 
	only degenerate scenarios and those with inverse mass hierarchy 
	survive. 
	This result allows neutrinos to still play 
	an important role as dark matter in the Universe.
	To improve the accuracy of the present result, considerably enlarged 
	experiments are required, such as GENIUS. 
	A GENIUS Test Facility has been funded and will 
	 come into operation by early 2003.}


\section{Introduction}

	Double beta decay is the most sensitive probe to test 
	lepton number conservation.
	Further it seems to be the only way to decide about 
	the Dirac or Majorana nature of the neutrino.
	Observation of $0\nu\beta\beta$ decay would prove that 
	the neutrino is a Majorana particle and would be another 
	clear sign of beyond standard model physics. 
	Recently atmospheric and solar neutrino oscillation experiments 
	have shown that neutrinos are massive. This was the first 
	indication of beyond standard model physics. The absolute 
	neutrino mass scale, however, cannot be determined 
	from oscillation experiments alone. 
	Double beta decay is indispensable also to solve this problem. 

	The observable of double beta decay is the effective neutrino mass

\centerline{$\langle m \rangle =
|\sum U^2_{ei}m^{}_i| = |m^{(1)}_{ee}| 
		      + e^{i\phi_2} |m^{(2)}_{ee}| 
		      + e^{i\phi_3} |m^{(3)}_{ee}|,
$}

\noindent
	  with $U^{}_{ei}$ 
	  denoting elements of the neutrino mixing matrix, 
	  $m_i$ neutrino mass eigenstates, and $\phi_i$  relative Majorana 
	  CP phases. It can be written in terms of oscillation parameters 
\cite{KKPS} 
\begin{eqnarray}
\label{1}
|m^{(1)}_{ee}| &=& |U^{}_{e1}|^2 m^{}_1,\\
\label{2}
|m^{(2)}_{ee}| &=& |U^{}_{e2}|^2 \sqrt{\Delta m^2_{21} + m^{2}_1},\\
\label{3}
|m^{(3)}_{ee}| &=& |U^{}_{e3}|^2 \sqrt{\Delta m^2_{32} 
				 + \Delta m^2_{21} + m^{2}_1}.
\end{eqnarray}

	The effective mass $\langle m \rangle$ is related with the 
	half-life for $0\nu\beta\beta$ decay via 
$\left(T^{0\nu}_{1/2}\right)^{-1}\sim \langle m_\nu \rangle^2$, 
        and for the limit on  $T^{0\nu}_{1/2}$
	deducible in an experiment we have 
\begin{eqnarray}
\label{4}	
T^{0\nu}_{1/2} \sim \epsilon \times a \sqrt{\frac{Mt}{\Delta E B}}, 
\end{eqnarray}

	Here $a$ is the isotopical abundance of the $\beta\beta$ emitter;
	$M$ is the active detector mass; 
	$t$ is the measuring time; 
	$\Delta E$ is the energy resolution; 
	$B$ is the background count rate 
	and $\epsilon$ is the efficiency for detecting a $\beta\beta$ signal. 
	Determination of the effective mass fixes the absolute 
	scale of the neutrino mass spectrum. 
\cite{KKPS,KK60Y}). 

	In this paper we will discuss the status of double 
	beta decay search. 
	We shall, in section 2, 
	discuss the 
	recent evidence for the neutrinoless decay mode, from the 
	HEIDELBERG-MOSCOW experiment   
	and the consequences for the neutrino mass scenarios 
	which could be realized in nature.
	In section 3 we discuss the 
	possible future potential of $0\nu\beta\beta$ experiments,
	which could improve the present accuracy. 
	



\begin{figure}[h]
\centering{
\includegraphics*[scale=0.45]
{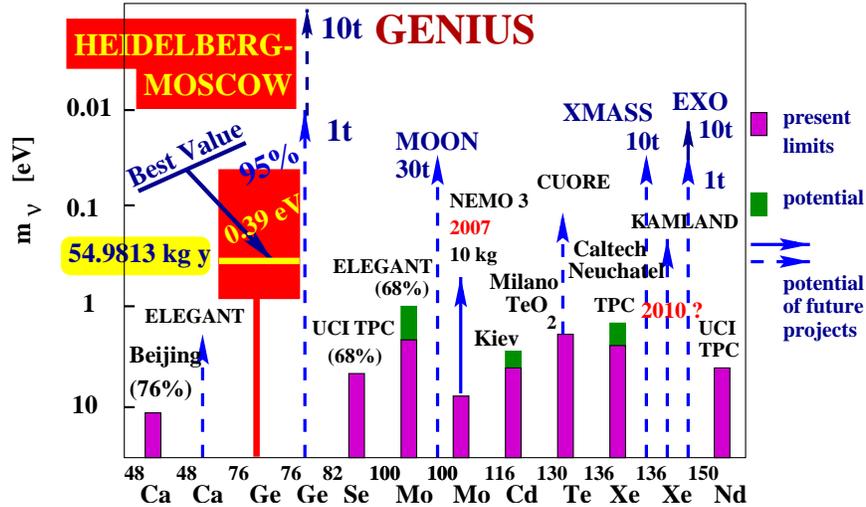}}
\caption[]{
       Present sensitivity, and expectation for the future, 
       of the most promising $\beta\beta$ experiments. 
       Given are limits for $\langle m \rangle $, except 
	for the HEIDELBERG-MOSCOW experiment where the recently 
	observed {\it value} is given (95$\%$ c.l. range and best value).
	Framed parts of the bars: present status; not framed parts: 
       future expectation for running experiments; solid and dashed lines: 
       experiments under construction or proposed, respectively. 
       For references see 
\protect\cite{KK60Y,KK02-PN,KK02-Found,KK-LowNu2,KK-NANPino00}.
\label{fig:Now4-gist-mass}}
\end{figure}



\section{Evidence for the Neutrinoless Decay Mode}

	The status of present double beta experiments is shown in 
Fig.~\ref{fig:Now4-gist-mass}	
	and is extensively discussed in 
\cite{KK60Y}.	
	The HEIDELBERG-MOSCOW experiment using the largest source strength 
	of 11 kg of enriched $^{76}$Ge (enrichment 86$\%$) 
	in form of five HP Ge-detectors 
	is running since August 1990 
	in the Gran-Sasso underground laboratory 
\cite{KK60Y,KK02-Found,HDM01,KK02,KK-StProc00,HDM97}, 
	and is since nine years now the most sensitive double 
	beta experiment worldwide. 


\subsection{\it Data from the HEIDELBERG-MOSCOW Experiment}

\begin{figure}[ht]
\begin{picture}(700,105)
\put( 5,147)
{\includegraphics{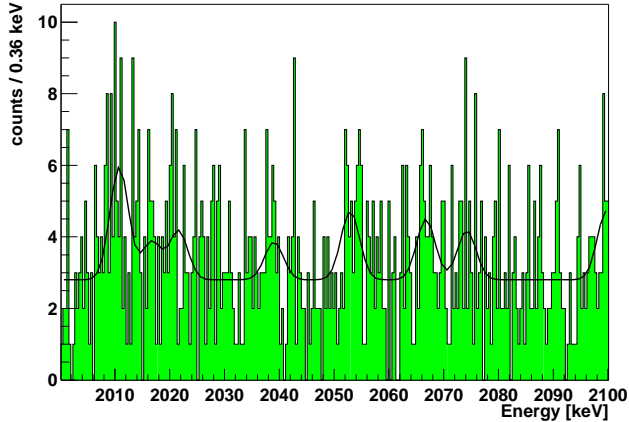}}
\end{picture}

\vspace{2.cm}
	\caption{The spectrum taken with the $^{76}{Ge}$ detectors 
	Nr. 1,2,3,4,5 over the period August 1990 - May 2000 
	(54.9813\,kg\,y) in the original 0.36\,keV binning, 
	in the energy range 2000 - 2100\,keV. Simultaneous 
	fit of the $^{214}{Bi}$ lines and the two high-energy 
	lines yield a probability for a line at 2039.0\,keV of 91\%.
\label{fig:100keV_sp}}
\end{figure}

	The data taken in the period August 1990 - May 2000 
	(54.9813\,kg\,y, or 723.44 mol-years) are shown in 
Fig.~\ref{fig:100keV_sp}
	in the section around the Q$_{\beta\beta}$ 
	value of 2039.006\,keV 
\cite{New-Q-2001,Old-Q-val}. 
Fig.~\ref{fig:100keV_sp}
	is identical with 
Fig.~\ref{fig:Now4-gist-mass} 
	in 
\cite{KK02}, 
	except that we show here the original energy binning 
	of the data of 0.36\,keV. These data have been analysed 
\cite{KK02,KK02-PN,KK02-Found}
	with various statistical methods, with the 
	Maximum Likelihood Method and 
	in particular also with the Bayesian method (see, e.g. 
\cite{Bayes_Method-General}).  
	This method is particularly suited for low
	counting rates, where the data follow 
	a Poisson distribution, that cannot be approximated by a Gaussian.
	Details and the results of the analysis are given in 
\cite{KK02,KK02-PN,KK02-Found}.	

\begin{figure}[h]
\begin{center}
\includegraphics[scale=0.2]
{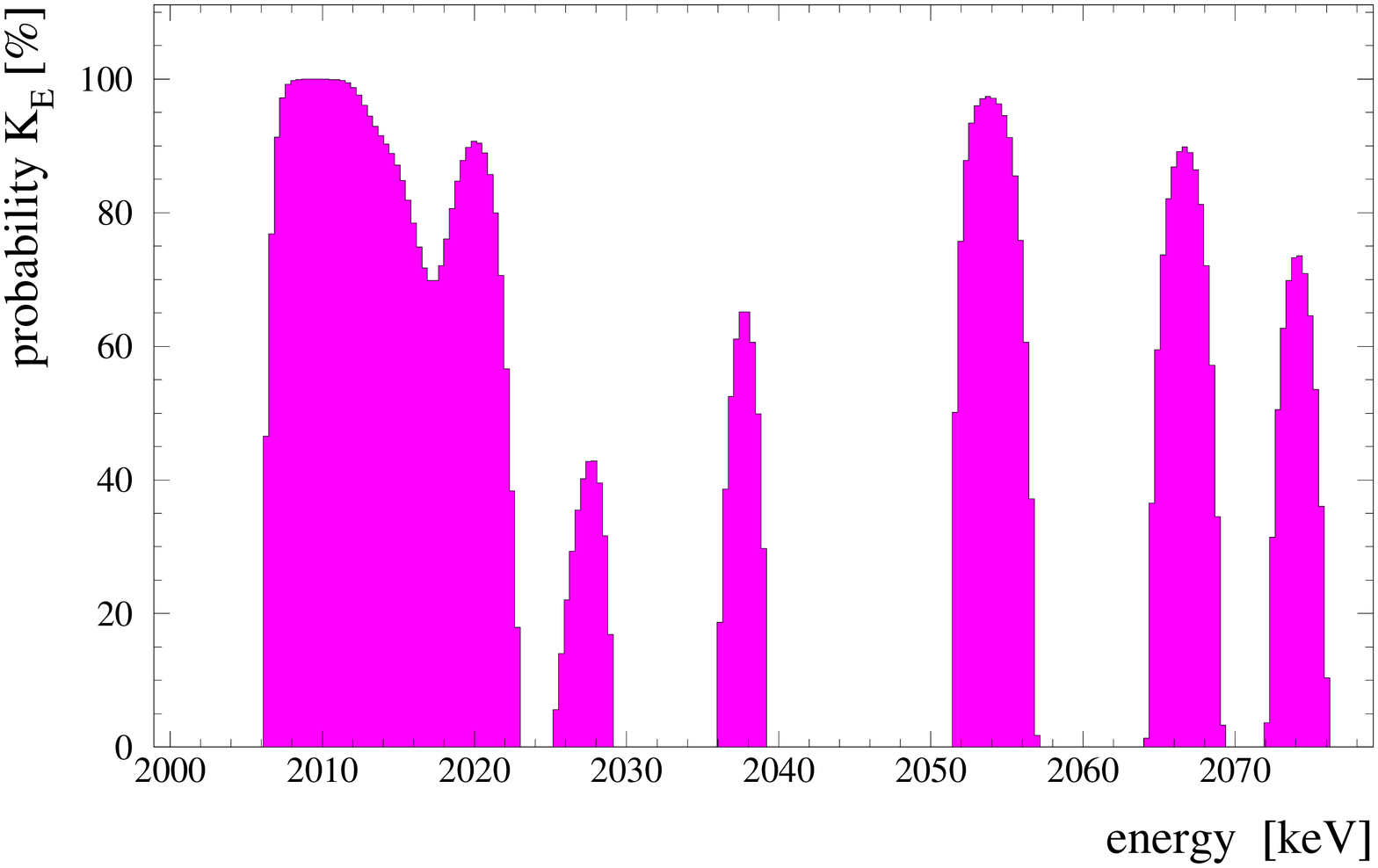} 
\hspace{.3cm}
\includegraphics[scale=0.2]
{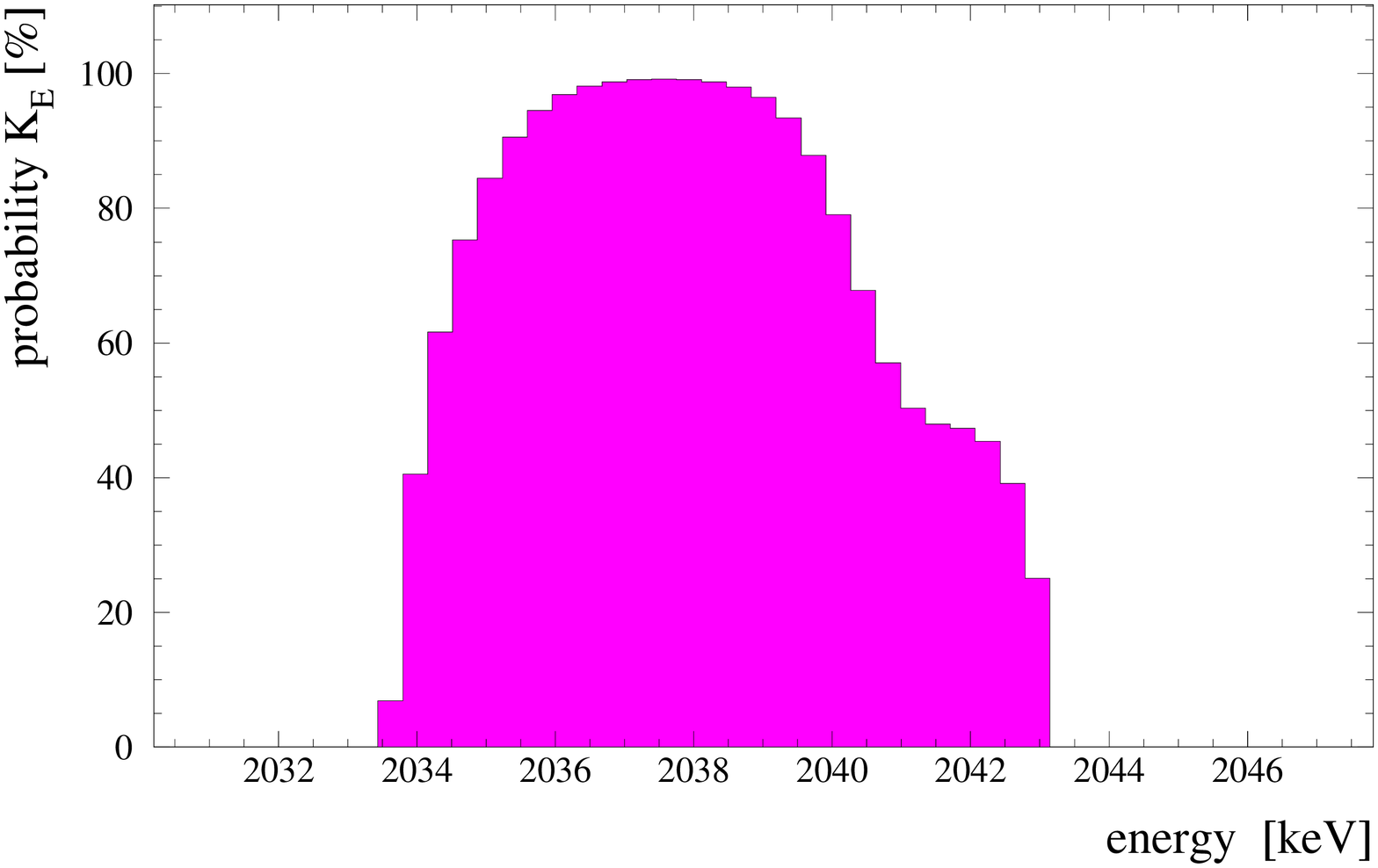} 
\end{center}

\vspace{-.7cm}
	\caption{Left: 
	Probability K that a line exists at a given energy in the 
	range of 2000-2080 keV derived via Bayesian inference 
	from the spectrum shown in Fig.
\protect\ref{fig:100keV_sp}. 
	Right: 
	Result of a Bayesian scan for lines as in 
	the left part of this figure,  
	but in an energy range of $\pm 5\sigma$ around Q$_{\beta\beta}$. 
\label{fig:Bay-Chi-all-90-00-gr}}
\end{figure}


	Our peak search procedure (for details see 
\cite{KK02-PN,KK02-Found}) 
	reproduces (see 
\cite{KK02,KK02-PN,KK02-Found}) 
	$\gamma$-lines 
	at the positions of  known weak lines  
	from the decay of $^{214}{Bi}$ at 2010.7, 2016.7, 2021.8 
	and 2052.9 keV 
\cite{Tabl-Isot96}. 
	In addition, a line centered at 2039 keV shows up 
(see Fig. \ref{fig:Bay-Chi-all-90-00-gr}). 
	This is compatible with the Q-value 
\cite{New-Q-2001,Old-Q-val}
	of the double beta decay process. The Bayesian analysis 
	yields, when analysing a $\pm5\sigma$ range around Q$_{\beta\beta}$ 
	(which is the usual procedure when searching for resonances 
	in high-energy physics)	a confidence level (i.e. the probability K) 
	for a line to exist at 
	2039.0 keV of 96.5 $\%$ c.l. (2.1 $\sigma$) 
(see Fig. \ref{fig:Bay-Chi-all-90-00-gr}).  
	We repeated the analysis for the same data, 
	but except detector 4, which had no muon shield 
	and a slightly worse energy resolution (46.502\,kg\,y). 
	The probability we find for a line at 2039.0 keV in this case 
	is 97.4$\%$ (2.2 $\sigma$) 
\cite{KK02,KK02-PN,KK02-Found}.

	Fitting a wide range of the spectrum yields a line 
	at 2039\,keV at 91\% c.l. (see 
Fig.\ref{fig:100keV_sp}). 

	We also applied the Feldman-Cousins method  
\cite{RPD00}.
	This method 
	(which does not use the information 
	that the line is Gaussian) finds 
	a line at 2039 keV on a confidence level of 
	3.1 $\sigma$ (99.8$\%$ c.l.). 
	In addition to the line at 2039\,keV we find candidates 
	for lines at energies beyond 2060\,keV and around 2030\,keV, 
	which at present cannot be attributed. This is a task 
	of nuclear spectroscopy. 
	
\begin{figure}[t]
\begin{center}
\includegraphics[scale=0.2]{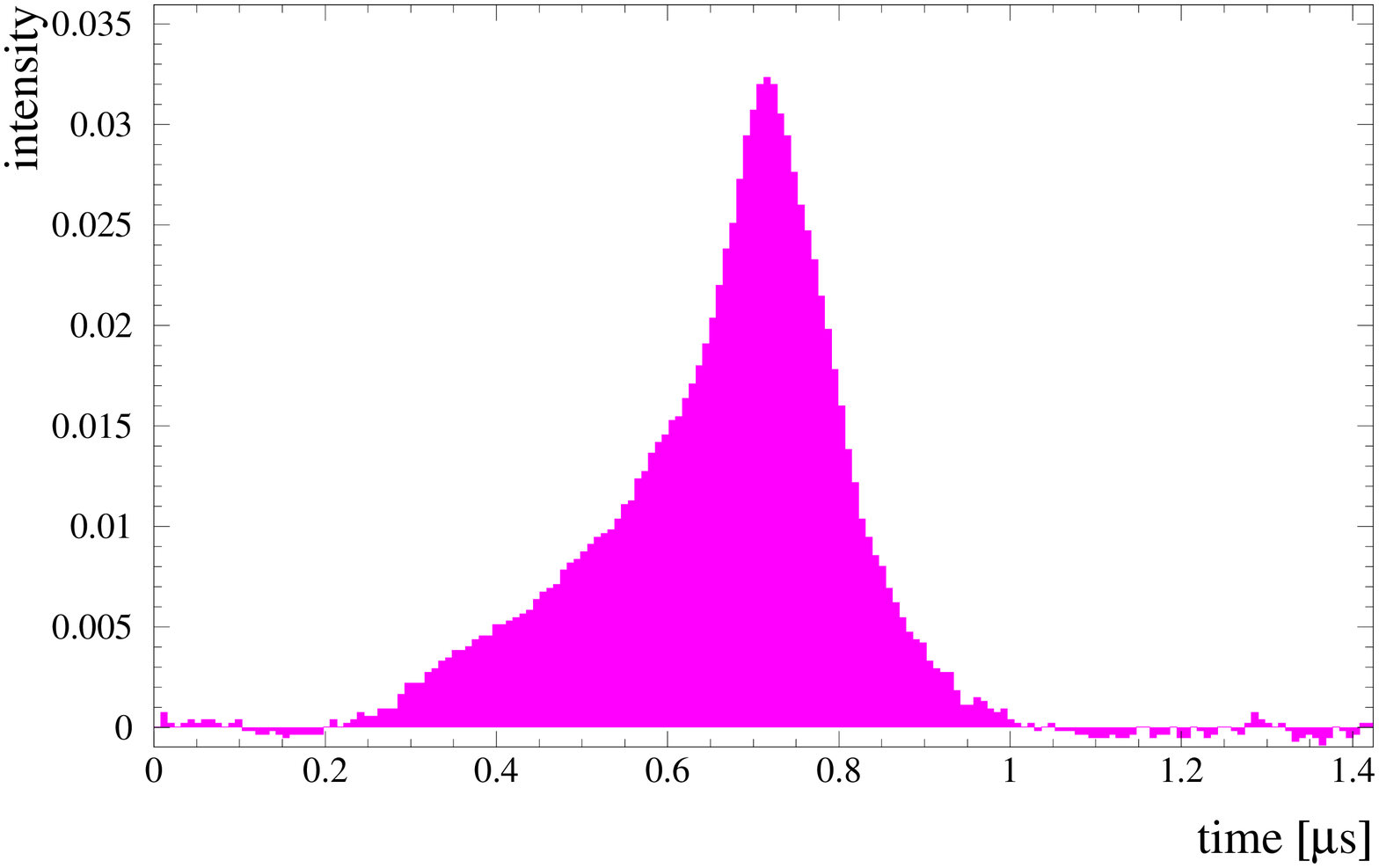} 
\hspace{.3cm}
\includegraphics[scale=0.2]{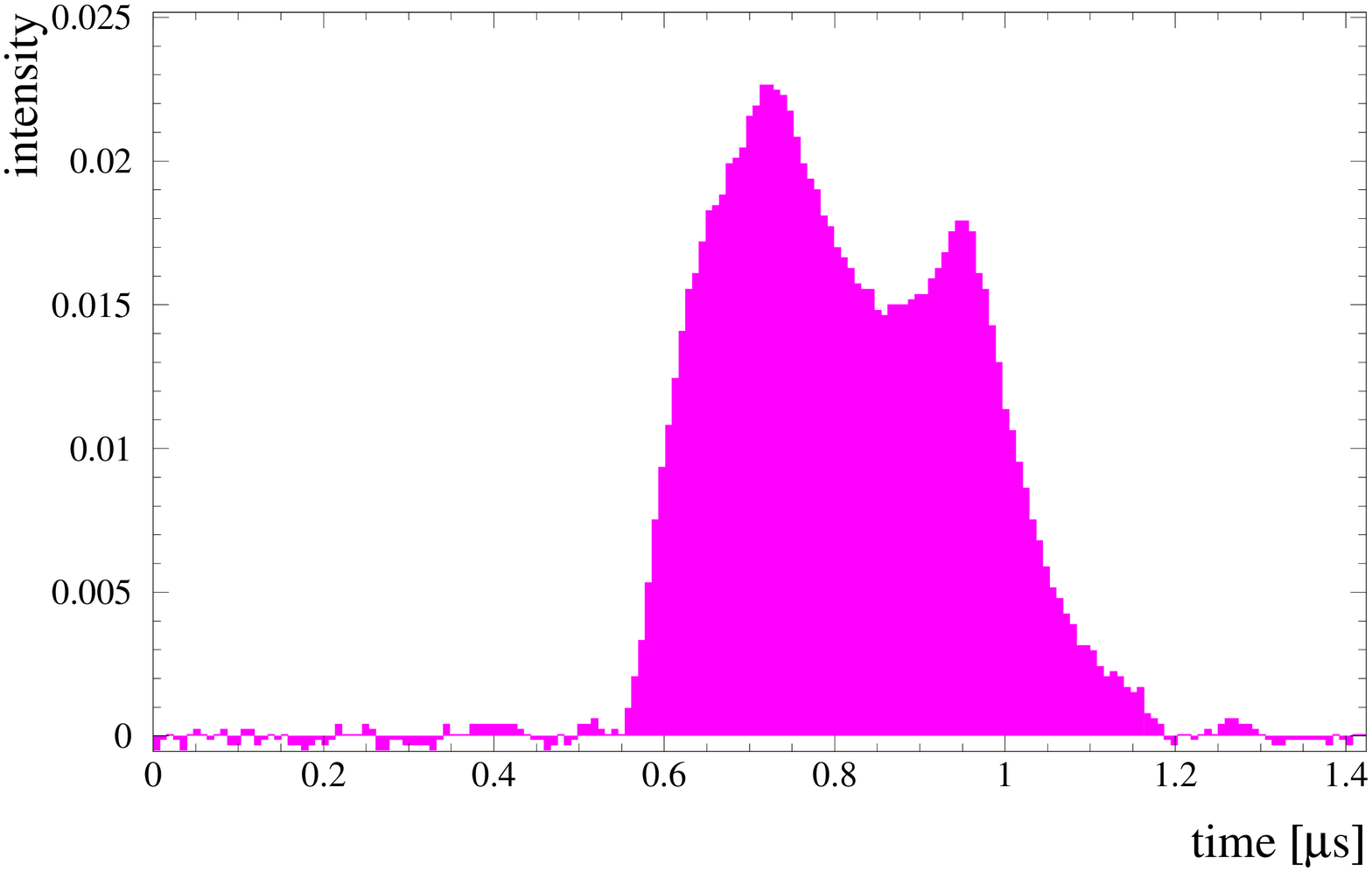} 
\end{center}

\vspace{-.7cm}
\caption{Left: Shape of one candidate for \znbb ~~decay  
	classified as SSE by all three methods 
	of pulse shape discrimination.  
	Right: Shape of one candidate 
	classified as MSE by all three methods.}
\label{fig:Shape}
\end{figure}

	Important further information can be obtained from the 
	{\it time structures} 
	of the individual events. 
	Double beta events should behave as single site events 
(see Fig. \ref{fig:Shape} left), 
	i.e. clearly different from a multiple scattered $\gamma$-event  
(see Fig. \ref{fig:Shape} right). 
	It is possible to differentiate between these different 
	types of events by pulse shape analysis. 
	We have developped three methods of pulse shape analysis 
\cite{HelKK00,Patent-KKHel,KKMaj99} 
	during the last seven years, one of which has been patented 
	and therefore only published recently. 	

	Installation of Pulse Shape Analysis (PSA) 
	has been performed in 1995 for the  
	four large detectors. Detector 
	Nr.5 runs since February 1995, detectors 2,3,4 since 
	November 1995 with PSA. 
	The measuring time with PSA  
	from November 1995 until May 2000 is 36.532 kg years, 
	for detectors 2,3,5 it is 28.053\,kg\,y.

\begin{figure}[h]

\vspace{-0.3cm}
\includegraphics[scale=0.2]
{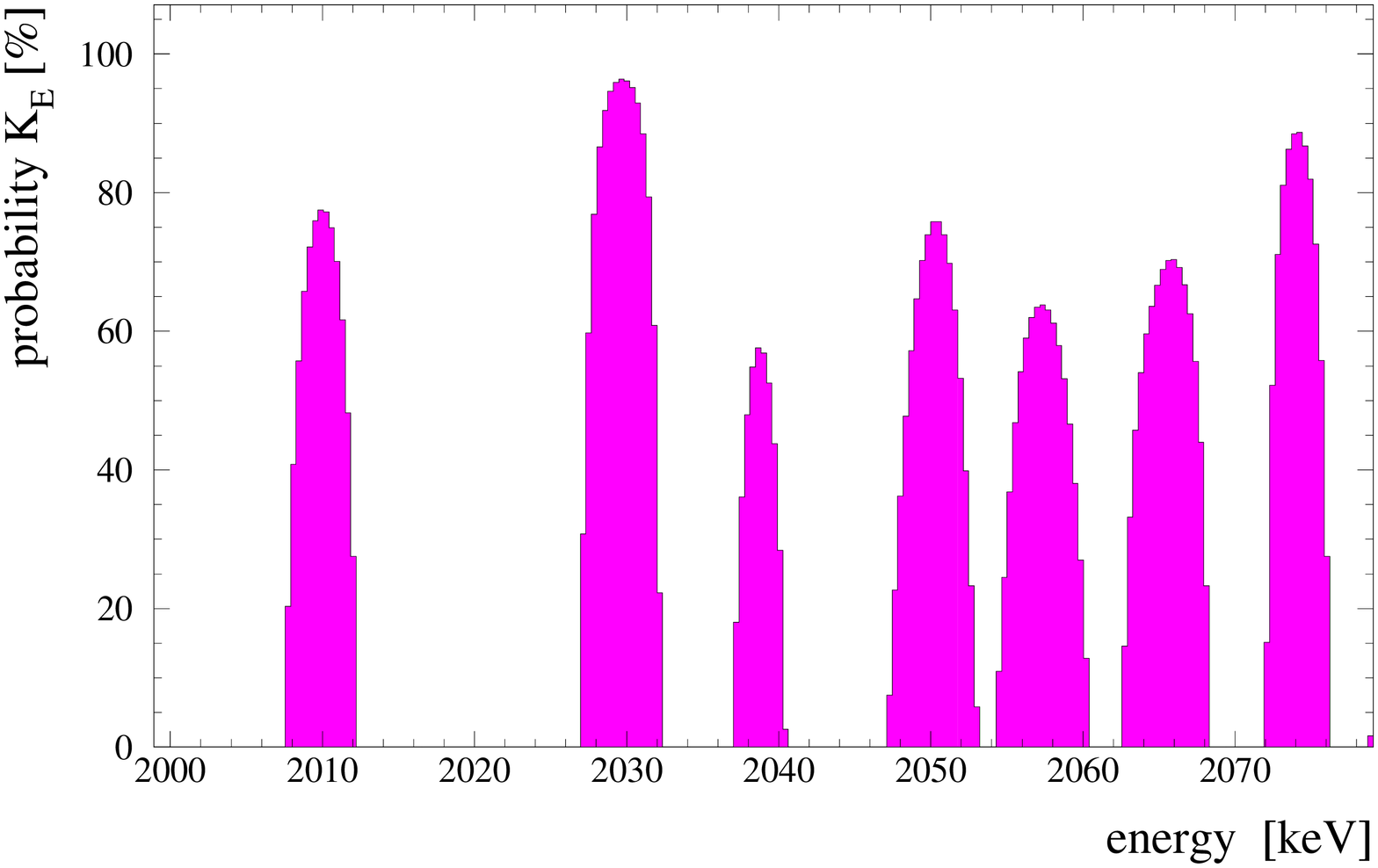} 
\hspace{0.3cm}
\includegraphics[scale=0.2]
{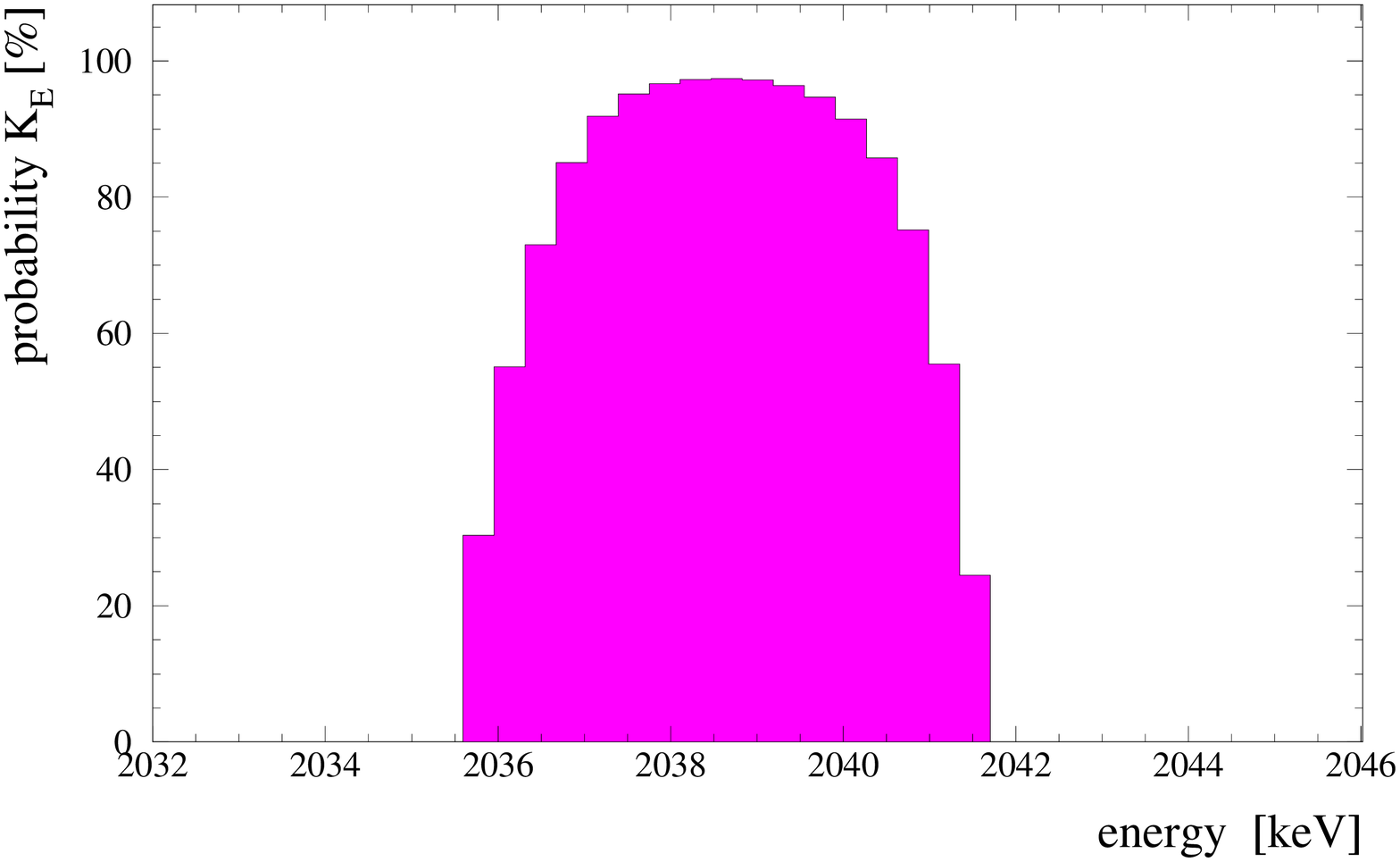} 

\vspace{-0.3cm}
\caption[]{Scan for lines in the single site event spectrum 
	taken from 1995-2000 with detectors Nr. 2,3,5, 
	(Fig. 
\ref{fig:Sum_spectr_5det}), 
	with the Bayesian method.
	\underline{Left:} Energy range 2000 -2080\,keV. 
	\underline{Right:} Energy range of analysis 
	$\pm\,4.4\sigma$ around Q$_{\beta\beta}$.} 
\label{fig:Bay-Hell-95-00-gr-kl-Bereich}


\begin{center}
\includegraphics[scale=0.33]
{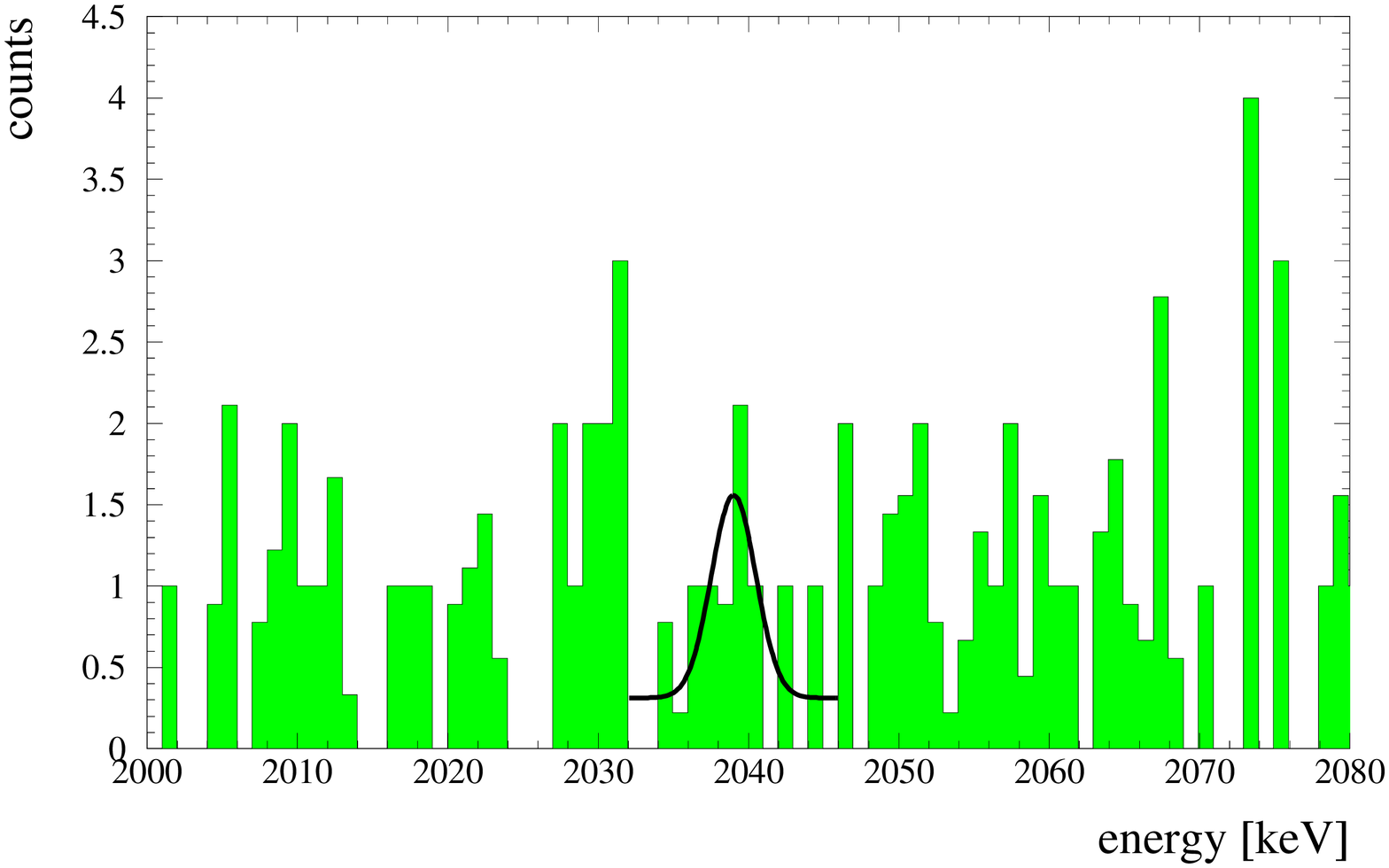} 
\end{center}

\vspace{-0.8cm}
\caption[]{Sum spectrum of single site events, 
	measured with the detectors 
	Nr. 2,3,5 operated with pulse shape analysis in the period 
	November 1995 to May 2000 (28.053\,kg\,y),  
	summed to 1\,keV bins. 
	Only events identified as single site events (SSE) 
	by all three pulse shape analysis methods 
\cite{HelKK00,Patent-KKHel,KKMaj99}
	have been accepted.
	The curve results from Bayesian inference in the 
	way explained in sec.3. 
	When corrected for the efficiency 
	of SSE identification (see text), 
	this leads to the following value for the half-life:
	T$_{1/2}^{0\nu}$=(0.88 - 22.38)$\times~10^{25}$~y (90\% c.l.).}
\label{fig:Sum_spectr_5det}
\end{figure}

	In the SSE spectrum obtained 
	under the restriction that the signal simultaneously fulfills  
	the criteria of {\it all three} methods for a single site event, 
	we find again indication of a line at 2039.0\,keV 
(see Figs. \ref{fig:Bay-Hell-95-00-gr-kl-Bereich}, \ref{fig:Sum_spectr_5det}).

	We find 9 SSE events in the 
	region 2034.1 - 2044.9\,keV 
	($\pm$ 3$\sigma$ around Q$_{\beta\beta}$).  
	Bayes analysis of the  range 2032 - 2046\,keV yields 
	a signal of single site events,  
	as expected 
	for neutrinoless double beta decay, with 96.8$\%$ c.l. 
	at the Q$_{\beta\beta}$ value.
	The Feldman-Cousins method gives a signal at 2039.0\,keV 
	of 2.8 $\sigma$ (99.4$\%$).


\begin{figure}[t]

\vspace{-.3cm}
\begin{center}
\includegraphics[scale=0.3]{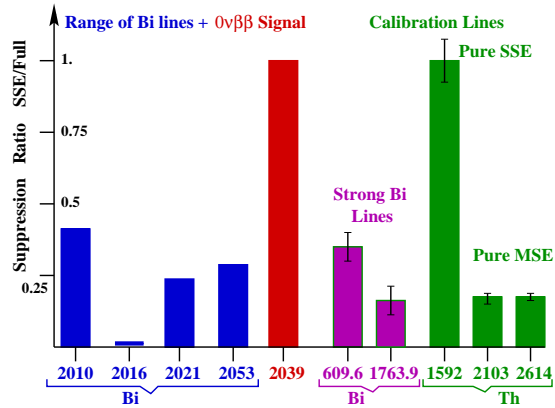} 
\end{center}

\vspace{-.3cm}
\caption{Relative suppression ratios: Remaining intensity 
	after pulse shape analysis compared to the intensity 
	in the full spectrum. Right: Result of a calibration 
	measurement with a Th source - ratio of the intensities 
	of the 1592\,keV line (double escape peak, known to be 100\% SSE), 
	set to 1. The intensities of the 2203\,keV line (single 
	escape peak, known to be 100\% MSE) are strongly reduced 
	(error bars are $\pm 1\sigma$. The same order of reduction 
	is found for the strong Bi lines occuring in our spectrum 
	- shown in this figure are the lines at 609.4 
	and 1763.9\,keV. Left: The lines in the range of weak 
	statistics around the line at 2039\,keV (shown are ratios 
	of best fit values). The Bi lines are reduced compared 
	to the line at 2039\,keV (set to 1), as to the 1592\,keV SSE Th line.}
\label{fig:Ratios}
\end{figure}

	The analysis of the line at 2039.0\,keV 
	before correction for the efficiency yields 
	4.6\,events (best value) or 
	(0.3 - 8.0)\,events within 95$\%$ c.l. 
	((2.1 - 6.8)\,events within 68.3$\%$ c.l.). 
	Corrected for the efficiency to identify 
	an SSE signal by successive application of all 
	three PSA methods, which is 0.55 $\pm$ 0.10, 
	we obtain a \znbb~ signal with 92.4$\%$ c.l.. 
	The signal is 
	(3.6 - 12.5)\,events with 68.3$\%$ c.l.
	 ~(best value 8.3\,events).
	Thus, with proper normalization concerning the running 
	times (kg\,y) of the full and the SSE spectra, 
	we see that almost the full signal remains after 
	the single site cut (best value), 
	while the $^{214}{Bi}$ lines (best values) are considerably reduced.
	We have used a $^{238}{Th}$ source to test the PSA method. 
	We find the reduction of the 2103\,keV 
	and 2614\,keV $^{228}{Th}$ lines (known to be multiple site 
	or mainly multiple site), relative to the 1592\,keV 
	$^{228}{Th}$ line (known to be single site), shown in Fig. 
\ref{fig:Ratios}. 
	This proves that the PSA method works efficiently. 
	Essentially the same reduction 
	as for the Th lines at 2103 and 2614\,keV and for the weak Bi lines 
	is found for the strong $^{214}{Bi}$ 
	lines (e.g. at 609.6 and 1763.9\,keV (Fig. 
\ref{fig:Ratios})).



	The possibility, that the single site signal is the double 
	escape line corresponding to a (much more intense!) 
	full energy peak of a $\gamma$-line, at 2039+1022=3061\,keV 
	is excluded from the high-energy part of our spectrum.


\begin{figure}[h]

\vspace{-1.cm}
\begin{center}
\includegraphics
[height=4.3cm,width=10.5cm]{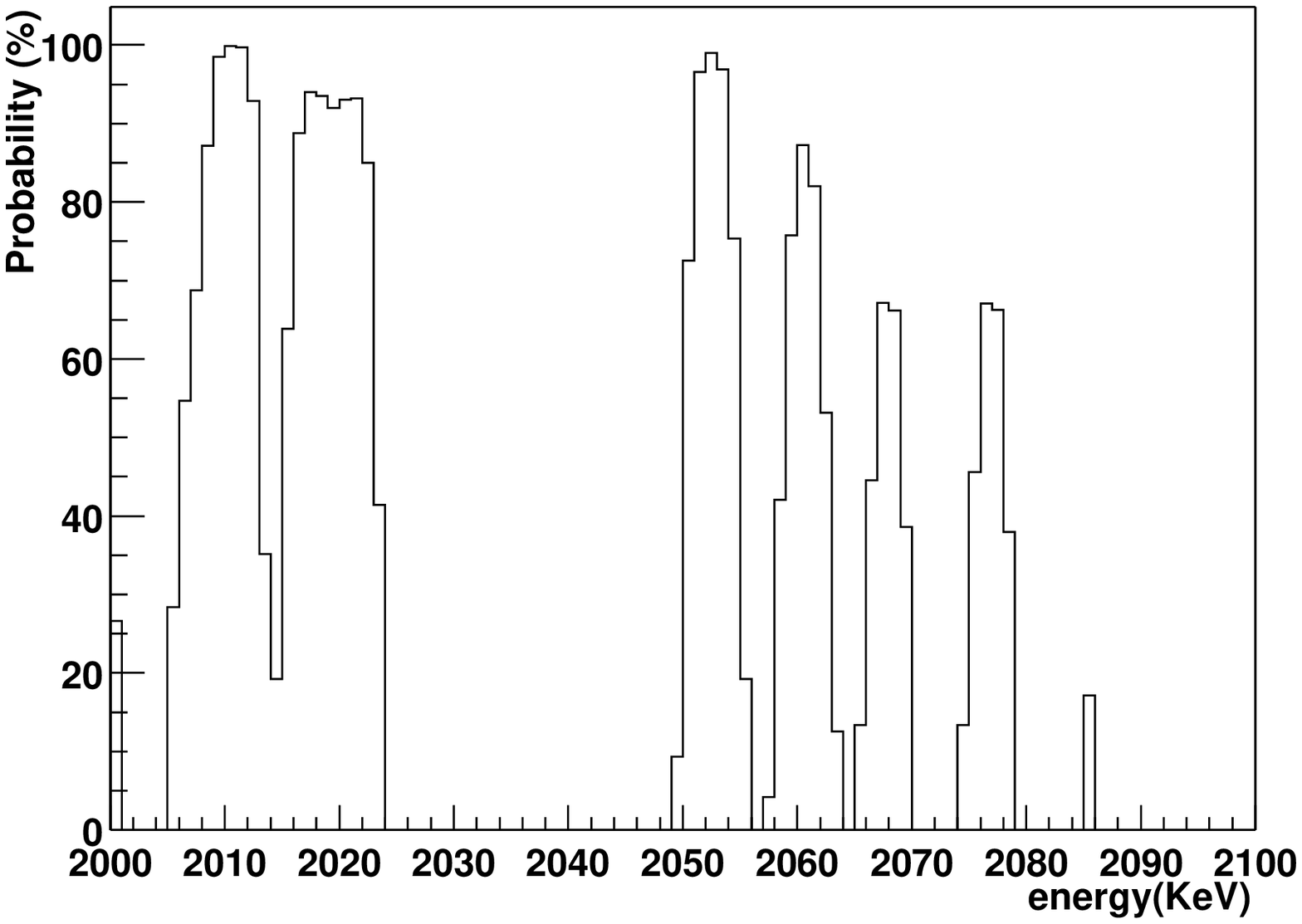}
\includegraphics
[height=4.3cm,width=10.5cm]{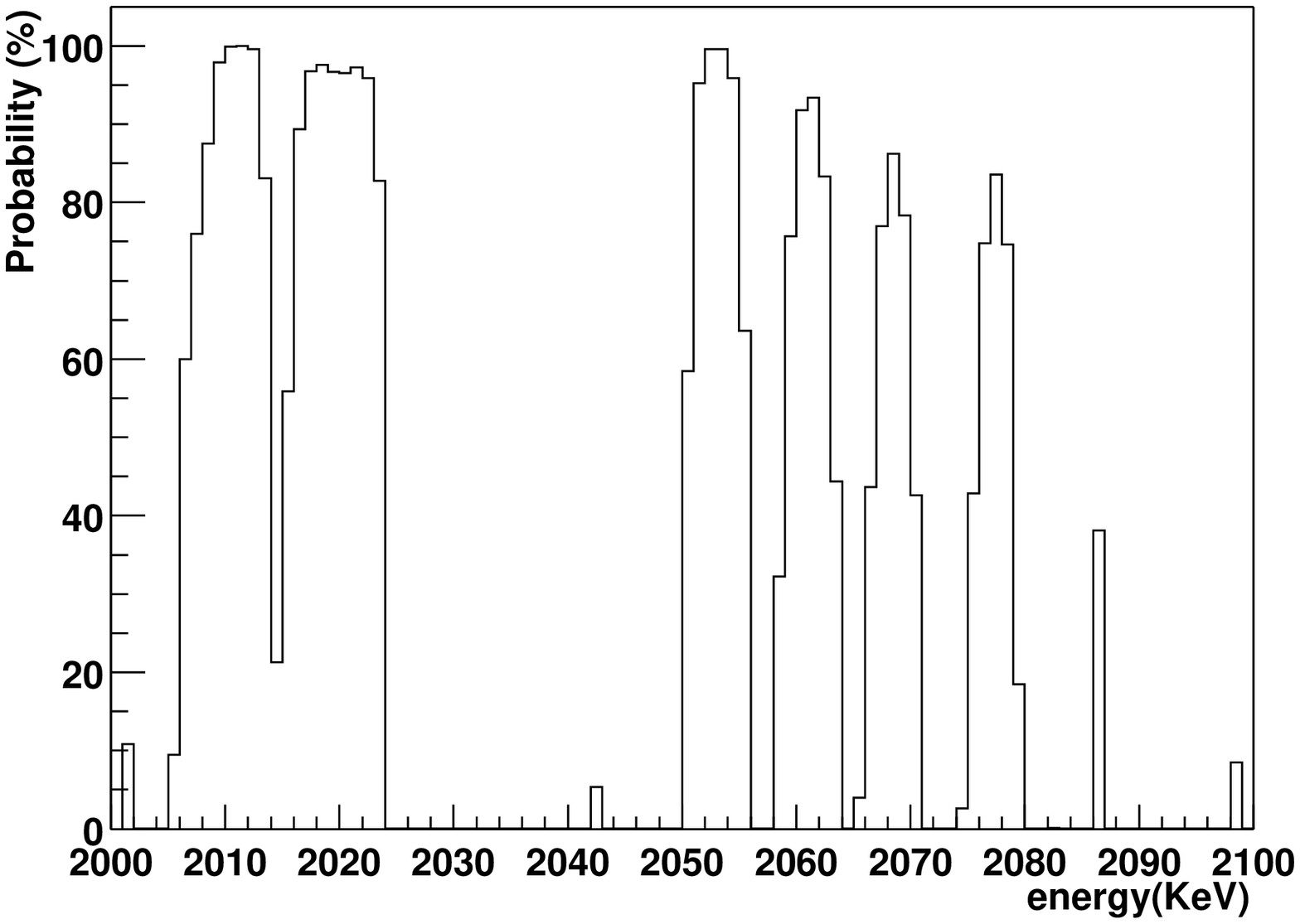}
\end{center}
\caption[]{Peak scanning of the spectrum measured by Caldwell et al. 
	\cite{Caldw91}, with the Maximum Likelihood method (upper part), 
	and with the Bayesian method (lower part) 
(as in Figs. 
\ref{fig:Bay-Chi-all-90-00-gr},\ref{fig:Sum_spectr_5det}) (see 
\cite{KK02-Found}).} 
\label{fig:Caldw-Max-Bayes}
\end{figure}
%
%

\subsection{Comparison with earlier results}

	We applied the same methods of peak search as used 
	in our analysis to the spectrum, measured 
	in the Ge experiment by Caldwell et al. 
\cite{Caldw91}
	more than a decade ago. 
	These authors had
	the most sensitive experiment using 
	{\it natural} Ge detectors (7.8\% abundance of $^{76}{Ge}$). 
	With their background being a factor of 9 higher than 
	in the present experiment, and their measuring time 
	of 22.6\,kg\,y, they have a statistics 
	for the background larger by a factor of almost 4 
	in their (very similar) experiment. 
	This allows helpful conclusions about the nature of the background.

	The peak scanning finds 
\cite{KK02-Found} 
	(Fig. 
\ref{fig:Caldw-Max-Bayes})
	indications for peaks essentially at  the same energies as in Fig.
 \ref{fig:Bay-Chi-all-90-00-gr}. 
	This shows that these peaks are not fluctuations. 
	In particular it sees 
	the 2010.78, 2016.7, 2021.6 and 2052.94\,keV $^{214}{Bi}$ 
	lines.
	It finds, however, {\it no line at Q$_{\beta\beta}$}.
	This is consistent with the expectation from the rate 
	found from the HEIDELBERG-MOSCOW experiment. 
	About 16 observed events in the latter correspond to 0.6 
	{\it expected} events in the Caldwell experiment, 
	because of the use of non-enriched material 
	and the shorter measuring time.

	Another Ge experiment (IGEX) using 8.8\,kg of enriched $^{76}{Ge}$, 
	but collecting since beginning of the experiment 
	in the early nineties till shutdown in end of 1999 only 8.8\,kg\,y 
	of statistics 
\cite{DUM-RES-AVIGN-2000}, 
	because of this low statistics also naturally cannot 
	see any signal at 2039\,keV. 
	This is consistent with our result. 
	The number of counts expected in that experiment 
	from our result would be about only 2.6. 
	In their spectrum however, which is unfortunately shown only 
	in the energy range 2020-2060\,keV, the line around 2030\,keV, 
	indicated in 
Figs. \ref{fig:Bay-Chi-all-90-00-gr},\ref{fig:Sum_spectr_5det},
	is also clearly seen.


\subsection{\it Proofs and Disproofs}

	The result described in section 2.1 has been questioned 
	in some papers 
	(Aalseth et al, hep-ex/0202018; Feruglio et al., Nucl. Phys. B
	637(2002)345; Zdesenko et al., Phys. Lett. B 546(2002) 206, and
	Kirpichnikov, talk at Meeting of Physical Section 
	of Russian Academy of Sciences, Moscow, December 2, 2002, 
	and priv. communication, Dec. 3, 2002.)
	These claims against our results are incorrect in various ways.

	The arguments in the first two of these papers 
	can be easily rejected.
	We have published this in hep-ph/0205228. 
	In particular their estimates of the intensities of the $^{214}{Bi}$ 
	lines are simply wrong, because the summing
	effect of the energies of consecutive gamma lines 
	(more precisely called True Coincidence Summing - TCS) 
	which is an elementary effect known in nuclear spectroscopy,  
	was not taken into account 
	(for a detailed discussion see 
\cite{KK02-Found}).
	Further none of the papers has performed a Monte
	Carlo simulation of our setup, which is the only way to come to
	quantitative statements.

	The paper by Zdesenko et al. starts from an arbitrary assumption, 
	namely that there are lines in the spectrum {\it at best} only 
	at 2010 and 2053\,keV.
	This contradicts to the experimental result, according 
	to which there are further lines in the spectrum 
	(see 
Figs. \ref{fig:Bay-Chi-all-90-00-gr},\ref{fig:Bay-Hell-95-00-gr-kl-Bereich},\ref{fig:Caldw-Max-Bayes}
	in the present paper).
	In this way they artificially increase the background 
	in their analysis and come to wrong conclusions.

	All three of these papers, when discussing the choice 
	of the width of the search window, ignore the results 
	of the statistical simulations we give
	in hep-ph/0205228 and we have published in 
\cite{KK02-PN} and \cite{KK02-Found}.

	Kirpichnikov states that from his analysis he clearly sees the
	2039\,keV line in the full (not pulse shape discriminated) 
	spectrum with $>99\%$ c.l. 
	He claims that he does not see the signal in the pulse shape
	spectrum. 
	One reason to see less intensity 
	certainly is that in this case he averages
	for determination of the background over 
	the full energy range without
	allowing for any lines. His result is in contradiction 
	with the result we obtain under the same assumption 
	of assuming just one line (at Q$_{\beta\beta}$)
	and a continuous background (see 
Fig. \ref{fig:Bay-Hell-95-00-gr-kl-Bereich} 
	of this paper).


\subsection{\it Half-Life and Effective Neutrino Mass}

	Having shown that the signal at Q$_{\beta\beta}$ 
	consists of single site events and is not a $\gamma$-line, 
	we translate the observed number of events into half-lifes. 
	We obtain 
\cite{KK02,KK02-PN,KK02-Found} 
	${\rm T}_{1/2}^{0\nu} = (0.8 - 18.3) \times 10^{25}$ 
	${\rm y}$ (95$\%$ c.l.) with a best value of $1.5 \times 10^{25}$y.
	Assuming that the $0\nu\beta\beta$ amplitude is dominated 
	by the neutrino mass mechanism, we obtain, with 
	the nuclear matrix element from 
\cite{Sta90} 
	an effective 
	mass of $\langle m_\nu \rangle$=(0.11 - 0.56)\,eV (95$\%$ c.l.).

	The result obtained is consistent
	 with all other double beta experiments -  
	which still reach less sensitivity. 
	The most sensitive experiments following the 
	HEIDELBERG-MOSCOW experiment are the geochemical $^{128}{Te}$ 
	experiment with 
	${\rm T}_{1/2}^{0\nu} > 2(7.7)\times 10^{24}
	{\rm~ y}$  (68\% c.l.), 
\cite{manuel}
	the $^{136}{Xe}$ experiment by the DAMA group with 
	${\rm T}_{1/2}^{0\nu} > 1.2 \times 10^{24}
	{\rm~ y}$  (90\% c.l.)%
\cite{DAMA02},  
	a second $^{76}{Ge}$ experiment with 
	${\rm T}_{1/2}^{0\nu} > 1.2 \times 10^{24}$ y
\cite{Kirpichn} 
	and a $^{nat}{Ge}$ experiment with 
	${\rm T}_{1/2}^{0\nu} > 1 \times 10^{24}$ y 
\cite{Caldw91,Kirpichn}.
	Other experiments are already about a factor of 100 
	less sensitive concerning the \znbb~ 
	half-life: the Gotthard TPC experiment with $^{136}{Xe}$ yields 
\cite{Gottch} 
	${\rm T}_{1/2}^{0\nu} > 4.4 \times 10^{23}
	{\rm~ y}$  (90\% c.l.) and the Milano Mibeta cryodetector experiment 
\cite{Ales00}
	${\rm T}_{1/2}^{0\nu} > 1.44 \times 10^{23}
	{\rm~ y}$  (90\% c.l.).

	Another expe\-riment 
\cite{DUM-RES-AVIGN-2000}
	with enriched $^{76}{Ge}$,
	which has stopped operation in 1999 after 
	reaching a significance of 8.8\,kg\,y,
	yields (if one believes their method of 'visual inspection' 
	in their data analysis), in a conservative analysis, 
	a limit of about 
	${\rm T}_{1/2}^{0\nu} > 5 \times 10^{24}
	{\rm~ y}$  (90\% c.l.). 
	The $^{128}{Te}$ geochemical experiment 
	yields $\langle m_\nu \rangle < 1.1$ eV (68 $\%$ c.l.)
\cite{manuel},   
	the DAMA $^{136}{Xe}$ experiment 
	$\langle m_\nu \rangle < (1.1-2.9)$\,eV 
\cite{DAMA02}
	and the $^{130}{Te}$ cryogenic experiment yields 
	$\langle m_\nu \rangle < 1.8$\,eV 
\cite{Ales00}. 

	Concluding we obtain, with about 95$\%$ probability, 
	first evidence for the neutrinoless 
	double beta decay mode. 
	As a consequence, at this confidence level, 
	lepton number is not conserved. 
	Further the neutrino is a Majorana particle. 
	If the 0$\nu\beta\beta$ amplitude is dominated by exchange 
	of a massive neutrino the effective mass 
	$\langle m \rangle $ is deduced 
	to be $\langle m \rangle $ 
	= (0.11 - 0.56)\,eV (95$\%$ c.l.), 
	with best value of 0.39\,eV. 
	Allowing conservatively for an uncertainty of the nuclear 
	matrix elements of $\pm$ 50$\%$
	(for detailed discussions of the status 
	of nuclear matrix elements we refer to 
\cite{KK60Y,KK02-Found,Tom91} 
	and references therein)
	this range may widen to 
	$\langle m \rangle $ 
	= (0.05 - 0.84)\,eV (95$\%$ c.l.). 

	Assuming other mechanisms to dominate the \znbb~ decay amplitude, 
	the result allows to set stringent limits on parameters of SUSY 
	models, leptoquarks, compositeness, masses of heavy neutrinos, 
	the right-handed W boson and possible violation of Lorentz 
	invariance and equivalence principle in the neutrino sector. 
	For a discussion and for references we refer to 
\cite{KK60Y,KK-Bey97,KK-Neutr98,KK-SprTracts00,KK-NANPino00}.

	With the limit deduced for the effective neutrino mass,  
	the HEIDELBERG-MOSCOW experiment excludes several 
	of the neutrino mass 
	scenarios 
	allowed from present neutrino oscillation experiments
	- allowing only for degenerate 
	and  inverse hierarchy mass scenarios 
\cite{KK-Sark01}. 

	Assuming the degenerate scenarios to be realized in nature 
	we fix - according to the formulae derived in 
\cite{KKPS} - 
	the common mass eigenvalue of the degenerate neutrinos 
	to m = (0.05 - 3.4)\,eV. 
	Part of the upper range is already excluded by 
	tritium experiments, which give a limit 
	of m $<$ 2.2-2.8\,eV (95$\%$ c.l.) 
\cite{Weinh-Neu00}.
	The full range can only  partly 
	(down to $\sim$ 0.5\,eV) be checked by future  
	tritium decay experiments,  
	but could be checked by some future $\beta\beta$

\begin{figure}[t]
\centering{
\includegraphics*[height=7.cm,width=12cm]
{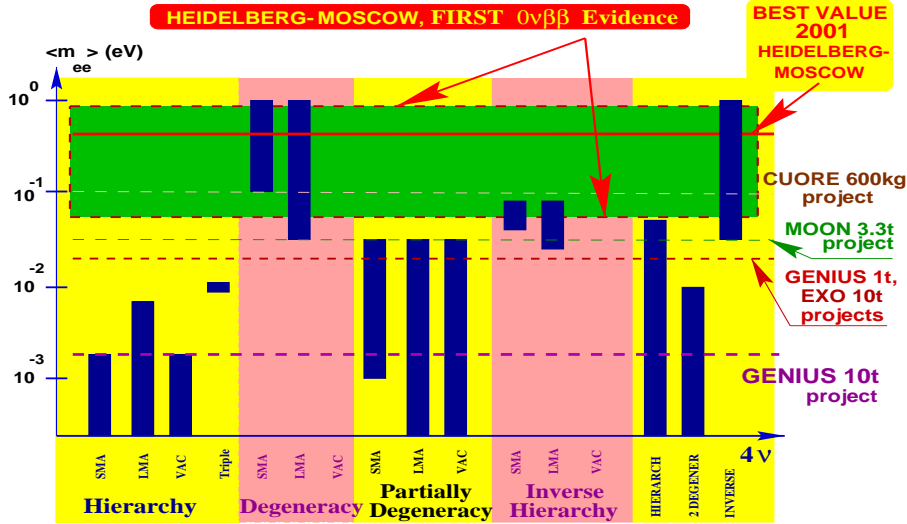}}
\caption[]{
	The impact of the evidence obtained for neutrinoless 
	double beta decay in this paper (best value 
	of the effective neutrino mass 
	$\langle m \rangle$ = 0.39\,eV, 95$\%$ 
	confidence range (0.05 - 0.84)\,eV - 
	allowing already for an uncertainty of the nuclear 
	matrix element of a factor of $\pm$ 50$\%$) 
	on possible neutrino mass schemes. 
	The bars denote allowed ranges of $\langle m \rangle$ 
	in different neutrino mass scenarios, 
	still allowed by neutrino oscillation experiments (see 
\cite{KK-Sark01}). 
	Hierarchical models are excluded by the 
	new \znbb ~~decay result. Also shown are 
	the expected sensitivities 
	for the future potential double beta experiments 
	CUORE, MOON, EXO  
	and the 1 ton and 10 ton project of GENIUS 
\cite{KK60Y,KK-SprTracts00,KK-00-NOON-NOW-NANP-Bey97-GEN-prop,GEN-prop}.
\label{fig:Jahr00-Sum-difSchemNeutr}}
\end{figure}
%
\begin{figure}[t]
\vspace{9pt}
\centering{
\includegraphics*[height=7.cm,width=12cm]
{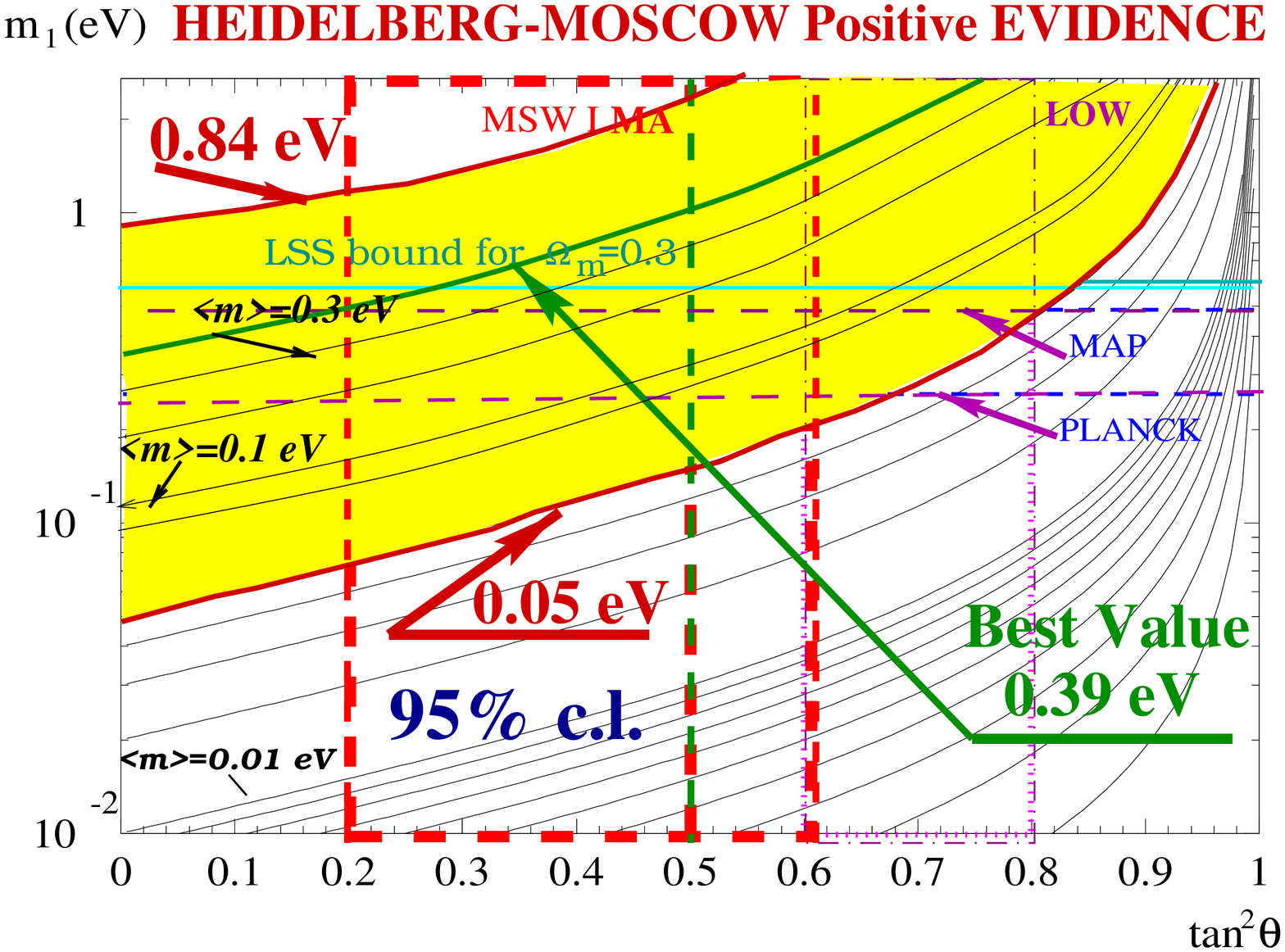}}

\vspace{-0.3cm}
\caption[]{
       Double beta decay observable 
$\langle m\rangle$
	 and oscillation parameters: 
	 The case for degenerate neutrinos. 
	 Plotted on the axes are the overall scale of neutrino masses 
	 $m_0$ and mixing $\tan^2\, \theta^{}_{12}$. 
	 Also shown is a cosmological bound deduced from a fit of 
	 CMB and large scale structure 
\cite{Lop} 
	and the expected sensitivity of the satellite experiments 
	 MAP and PLANCK. 
	 The present limit from tritium $\beta$ decay of 2.2\,eV 
\cite{Weinh-Neu00} 
	would lie near the top of the figure. 
     The range of 
$\langle m\rangle$
	 fixed by the HEIDELBERG-MOSCOW experiment 
\cite{KK02}
	is, in the case of small solar neutrino mixing, already in the 
	 range to be explored by MAP and PLANCK 
\cite{Lop}.
\label{fig:Dark3}}
\end{figure}


\noindent
	experiments (see, e.g., next section).
	The deduced best value for the mass 
	is consistent with expectations from experimental 
	$\mu ~\to~ e\gamma$
	branching limits in models assuming the generating 
	mechanism for the neutrino mass to be also responsible 
	for the recent indication for as anomalous magnetic moment 
	of the muon
\cite{MaRaid01}.  
	It lies in a range of interest also for Z-burst models recently 
	discussed as explanation for super-high energy cosmic ray events 
	beyond the GKZ-cutoff 
\cite{Farj00-04keV,PW01-Wail99,Fod-Bey02}. 
	A recent model with underlying A$_4$ symmetry for 
	the neutrino mixing matrix also leads to degenerate 
	neutrino masses consistent with the present result 
	from \znbb~ decay 
\cite{BMV02}. 
	The range of $\langle m \rangle $ fixed in this 
	work is, already now, in the range to be explored 
	by the satellite experiments MAP and PLANCK 
\cite{Lop,KKPS} 
	(see Fig. 
\ref{fig:Dark3}).

	The neutrino mass deduced leads to 0.002$\ge \Omega_\nu h^2 \le$
	0.1 and thus may allow neutrinos to still play 
	an important role as hot dark matter in the Universe 
\cite{KK-LP01}.


\section{Future of $\beta\beta$ Experiments - GENIUS and Other Proposals}

	With the HEIDELBERG-MOSCOW experiment, the era of the small smart 
	experiments is over. 
	New approaches and considerably enlarged experiments 
	(as discussed, e.g. in 
\cite{KK60Y,KK-Neutr98,KK-00-NOON-NOW-NANP-Bey97-GEN-prop,GEN-prop,KK-NOW00})
	will be required in future 
	to fix the neutrino mass with higher accuracy. 
	
	Since it was realized in the HEIDELBERG-MOSCOW experiment, 
	that the remaining small background is coming from the material 
	close to the detector (holder, copper cap, ...), 
	elimination of {\it any} material close to the detector 
	will be decisive. Experiments which do not take this 
	into account, like, e.g. CUORE 
\cite{Cuore01},
	and MAJORANA 
\cite{MAJOR-WIPP00},
	will allow at best only rather limited steps in sensitivity. 
	Furthermore there is the problem in cryodetectors that they 
	cannot differentiate between a $\beta$ and a $\gamma$ signal, 
	as this is possible in Ge experiments.

	Another crucial point is - see eq. (4) - the energy resolution, 
	which can be optimized {\it only} in experiments 
	using Germanium detectors or bolometers. 
	It will be difficult to probe evidence for this rare decay 
	mode in experiments, which have to work - as result of their 
	limited resolution - with energy windows around 
	Q$_{\beta\beta}$ of several hundreds of keV, such as NEMO III%
\cite{ApPec02}, 
	EXO%
\cite{EXO-LowNu2}, 
	CAMEO%
\cite{CAMEO-Taup01}.

	Another important point is (see eq. 4), the efficiency 
	of a detector for detection of a $\beta\beta$ signal.
	For example, with 14$\%$ efficiency a potential 
	future 100\,kg $^{82}{Se}$ NEMO experiment would be, because 
	of its low efficiency, equivalent only to a 10\,kg 
	experiment (not talking about the energy resolution).

	In the first proposal for a third generation double 
	beta experiment, the GENIUS proposal 
\cite{KK-Bey97,KK-H-H-97,KK-Neutr98,KK-00-NOON-NOW-NANP-Bey97-GEN-prop,GEN-prop},
	the idea is to use 'naked' Germanium detectors in a huge tank 
	of liquid nitrogen. It seems to be at present the {\it only} 
	proposal, which can fulfill {\it both} requirements 
	mentioned above - to increase the detector mass 
	and simultaneously reduce the background drastically. 
	GENIUS would - with only 100\,kg of enriched $^{76}{Ge}$ - 
	increase the confidence level of the present pulse shape 
	discriminated 0$\nu\beta\beta$ signal to 4$\sigma$ within 
	one year, and to 7$\sigma$ within three years of measurement 
	(a confirmation on a 4$\sigma$ level by the MAJORANA project 
	would at least need $\sim$230\,years, the CUORE project would 
	need (ignoring for the moment the problem of identification 
	of the signal as a $\beta\beta$ signal) 3700 years).
	With ten tons of enriched $^{76}{Ge}$ GENIUS should be capable 
	to investigate also whether the neutrino mass mechanism 
	or another mechanism (see, e.g. 
\cite{KK60Y})
	is dominating the \znbb~ decay amplitude.
	A GENIUS Test Facility is at present under construction 
	in the GRAN SASSO Underground Laboratory 
\cite{GenTF-0012022,KK-GeTF-MPI}.

\vspace{-0.3cm}
\section{Conclusion}

	The status of present double beta decay search 
	has been discussed, and 	
	recent evidence for a non-vanishing 
	Majorana neutrino mass obtained by the HEIDELBERG-MOSCOW 
	experiment has been presented. Future projects to improve 
	the present accuracy of the effective neutrino mass have 
	been briefly discussed. The most sensitive of them and perhaps  
	at the same time most realistic one, is the GENIUS project.
	 GENIUS is the only of the new projects 
	which simultaneously has a huge potential for 
      cold dark matter search, and for real-time detection of 
      low-energy neutrinos (see 
\cite{KK-Bey97,KK-NOW00,BedKK-01,KK-SprTracts00,KK-IK,KK-LowNu2,KK-NANPino00}).

\section{Acknowledgment}

	The author is indebted to  his colleaques A. Dietz 
	and I.V. Krivosheina for the fruitful and pleasant collaboration, 
	and to C. Tomei and C. D\"orr 
	for their help in the analysis of the Bi lines. 
	The generous hospitality by Profs. D. Ahluwalia 
	and M. Kirchbach during the Zacatecas Forum on Physics 
	is gratefully acknowleged.


        

\end{document}